\input{aipcheck}

\documentclass[
    ,final            
  ]
  {aipproc}

\layoutstyle{8x11double}

\newcommand\aipcs{{Am.~Inst.~Phys.~Conf.~Ser.}}%
\newcommand\araa{{ARA\&A}}%
\newcommand\apj{{ApJ}}%
\newcommand\apjl{{ApJ}}%
\newcommand\apjs{{ApJS}}%
\newcommand\aap{{A\&A}}%
\newcommand\mnras{{MNRAS}}%
\newcommand\ssr{{Space~Sci.~Rev.}}%
\newcommand\nat{{Nature}}%
\newcommand\jgr{{J.~Geophys.~Res.}}%
\newcommand{\commentpcf}[1]{}

\def\ebv{E(B-V)}

\def\HI{H$^\mathrm{o}$}
\def\NHI{N(H$^\mathrm{o}$)}

\def\DI{D$^\mathrm{o}$}
\def\HeI{He$^\mathrm{o}$}
\def\FeII{Fe$^\mathrm{+}$}
\def\MgII{Mg$^\mathrm{+}$}
\def\cc{cm$^{-3}$}
\def\cmtwo{cm$^{-2}$}
\def\kms{km s$^{-1}$}
\def\mG{$\mu$G}
\def\deeg{$^{\circ}$}
\def\glon{$\ell$}
\def\glat{$b$}
\def\elon{$\lambda$}
\def\elat{$\beta$}
\def\glat{$b$}

\begin{document}

\title{The Heliosphere---Blowing in the Interstellar Wind}


\author{Priscilla C. Frisch}{
  address={University of Chicago, Chicago, IL 60637 USA}
}

\begin{abstract}
Measurements of the velocity of interstellar \HeI\ inside of the
heliosphere have been conducted over the past forty years.  These
historical data suggest that the ecliptic longitude of the direction
of the interstellar flow has increased at an average rate of $\sim
0.19^\circ$ per year over time.  Possible astronomical explanations
for these short-term variations in the interstellar gas entering the
heliosphere are presented.
\end{abstract}

\maketitle


\section{Introduction}

Our heliosphere is 'blowing in the wind' of the surrounding Local
Interstellar Cloud (LIC). Like all winds, the interstellar medium
(ISM) around the heliosphere changes with time over both short and
long timescales. These variations cause changes in the flux of
galactic cosmic rays at 1 AU that may imprint on the radionuclide
record on Earth \cite{FrischMueller:2011ssr}.

When Leverett Davis, Jr. first proposed that the solar wind carves
out a cavity in the galactic magnetic field \cite{Davis:1955}, he
noted that the path taken by galactic cosmic rays (GCRs) to the inner
heliosphere depends on the directions between the solar and
interstellar magnetic fields (ISMF).  Although the angle between the
ISMF direction and the motion of interstellar gas flowing
through the heliosphere was recognized as important for determining
the termination shock distance
\cite{Holzer:1989}, the magnetically distorted heliosphere was confirmed
only after Voyager 2 crossed the solar wind termination shock in 2008
\cite{Stoneetal:2008nature}. 

Interstellar gas and dust have now been measured in the inner
heliosphere by many spacecraft
\cite{Frisch:2011araa}.  The direction of the local ISMF has been
independently determined from the geometry of the band, or 'Ribbon', of 
energetic neutral atoms (ENAs)
discovered by the Interstellar Boundary Explorer (IBEX) spacecraft
\cite{McComas:2009sci}.  The modulation of galactic cosmic rays
by the heliosphere has been known since Davis first proposed its
existence, but only recently have asymmetries in the GeV-TeV GCRs been
directly attributed to the modulation or acceleration of GCRs in the
plasma heliosphere
\cite{LazarianDesiati:2010,Schwadron:2012gcr}.

During the forty years over which the wind of interstellar \HeI\
through the heliosphere has been measured, the Sun has moved 160 AU
through space, and 200 AU with respect to the LIC.  Someday we might
encounter one of the tiny AU-sized low column density interstellar
``clouds'', \NHI$<10^{18} - 10^{19}$ \cmtwo, found throughout the ISM
\cite{GALFA:2012sss}. The nearest cloud  that has been identified of this
 type is in the constellation of Leo, $\sim 20$ pc away, where
interstellar absorption lines reveal cold (20 K), dense (3000 \cc),
and tiny clouds (200 AU)
\cite{Meyer:2012sss}.

In this paper, our galactic environment is briefly summarized and
evidence for short-term variations in the ISM feeding \HeI\ into the
heliosphere is presented, together with the implications of these
variations for the local interstellar environment.

\section{The interstellar magnetic field near the heliosphere}

Interstellar magnetic fields are difficult to measure over small
spatial scales in local ISM where densities are low, $n \sim
0.26$ \cc.  A diagnostic of the very local ISMF direction was
unexpectedly provided by the IBEX discovery of a 'Ribbon' of
ENAs \cite{McComas:2009sci}.  Agreement between the outer heliosphere
boundary conditions predicted by MHD heliosphere models
\cite{Pogorelovetal:2009ibexribbonasymmetries} and photoionization
models of the LIC \cite{SlavinFrisch:2008} has shown that the
heliosphere is located in a partially ionized interstellar cloud with
fairly well defined interstellar boundary conditions.  The heliosphere
models also predict the direction of ISMF in the outer heliosheath.
The directions traced by the 1 keV ENAs in the Ribbon are found to be
aligned with the model predictions for the ISMF direction tens of AU
outside of the heliopause. This coincidence suggests that the Ribbon
traces sightlines where the ISMF draping over the heliosphere is
perpendicular to the radial sightline
\cite{Schwadron:2009sci}.  Although the detailed 
physics of the Ribbon formation mechanism is not yet modeled, the
direction of the ISMF that shapes the heliosphere is thought to be at
the center of the 1 keV Ribbon arc.  A local ISMF strength of $\sim 3$
\mG\ is needed to equilibrate the ENA pressure in the inner
heliosheath with the total interstellar pressure
\cite{Schwadronetal:2011sep}.  This agrees with the field strength
derived from equality of thermal and magnetic pressure in the LIC gas
\cite{SlavinFrisch:2008,Frisch:2012howlocal}.

The geometrical structure of the Ribbon is reproduced by MHD
heliosphere models that suggest the Ribbon originates 10--100 AU
upstream of the heliopause
\cite{Schwadron:2009sci,Heerikhuisen:2010ribbon,HeerikhuisenPogorelov:2011,Ratkiewicz:2012ribbon,Chalov:2010ribbon}.  
In the models, the Ribbon is displaced towards the equator of the
undistorted ISMF (which corresponds to sightlines perpendicular to the
distant ISMF direction) as either the field strength becomes stronger
\cite{HeerikhuisenPogorelov:2011,Ratkiewicz:2012ribbon}, or as the
ribbon formation region moves further from the heliopause
\cite{FrischMcComas:2010}.  

Further from the Sun and within 40 pc, the ISMF direction has been
determined from measurements of the polarization of starlight caused
by magnetically aligned dust grains in nearby interstellar clouds
\cite{Frisch:2012ismf2}.  The polarization vector is expected to be
parallel to the ISMF direction in the diffuse ISM
\cite{Andersson:2012rev}.  We have developed a method to determine the
ISMF direction from an ensemble of polarization position angle data.
\footnote{The polarization position angle is defined as the angle
between the polarization direction and the north pole.}  
We have found that the nearest ISMF is aligned with the direction of
\glon,\glat$=47^\circ \pm 20^\circ, 25^\circ \pm 20^\circ$.
This direction is $32^\circ \pm 30^\circ$ from the center of the
Ribbon arc, located at \glon,\glat$=33^\circ \pm 4^\circ, 55^\circ
\pm 4^\circ$.  
Neither the optical
polarization data nor the Ribbon provide the polarity of the ISMF.

\begin{table}[h!]
\caption{Interstellar gas velocity obtained from interstellar \HeI\ data } \label{tab:one}
\begin{tabular}{ccccc}
\hline \tablehead{1}{c}{b}{Start/End\\of data\tablenote{Column 1 shows
the first and last year of data acquisition.}}
  & \tablehead{1}{c}{b}{Velocity\\(\kms)} 
   & \tablehead{1}{c}{b}{Downwind\\ \elon,\elat \tablenote{The
 uncertainties enclosed in brackets are assumed.}} 
  & \tablehead{1}{c}{b}{Coord.\\epoch\tablenote{For the pre-1990 and Ulysses \cite{Witte:2004} data,
the J2000 coordinates were
obtained by precessing B1950 celestial coordinates.}}
  & \tablehead{1}{c}{b}{Ref.} \\ 
\hline 

1972.8/1973.6 & 5--20 & $72.9,~-7.4~(\pm 3) $& 2000 & Weller \& Meyer \cite{WellerMeier:1974} \\ 

1974.1/1974.1 & [$22 \pm 3$]\tablenote{The \emph{Copernicus} velocity was assumed from Adams \& Frisch \cite{AdamsFrisch:1977}.}  
& $73.1,~-5.4~(\pm 5) $ & 2000 & Ajello et al. \cite{Ajello:1979}\\ 

1975.6/1976.1 & 9--27 &  $72.8,~-7.4~[\pm 3]$ & 2000 & Weller \& Meier\cite{WellerMeier:1979}\\   
1976.9/1977.1 & 22--28  & $73.2,~-4.4~[\pm 3]$ & 2000 &  Weller \& Meier \cite{WellerMeier:1981}\\

1977.8/1978.1 & $27 \pm 3$ &   $74.8 \pm 3,~-6.0 \pm 3$ & 2000 &  Dalaudier et al. \cite{DalaudierBertaux:1984}\\

1990.9/2002.7 &  $26.3 \pm 0.4$  & $75.4 \pm 0.5,~ -5.2 \pm 0.2$ & 2000 & Witte \cite{Witte:2004}\\

1992.6/1993.6 & $26.4 \pm 1.5 $    &   $76.0 \pm 0.4,~ -5.4 \pm 0.6$ &    & Flynn et al. \cite{FlynnVallerga:1998} \\
1992.6/2000.9 & $24.5 \pm 2.0 $    &   $74.7 \pm 0.5,~ -5.7 \pm 0.5$ &   & Vallerga et al. \cite{Vallergaetal:2004} \\

2009.2/2010.2 & $23.2 \pm 0.4$ & $79.00 \pm 0.47,~ -4.98 \pm 0.21$ & 2000 & McComas et al. \cite{McComas:2012bow} \\
\hline
\end{tabular}
\end{table}



 The ISMF traced by the polarization data appears to have an ordered
component that extends to within 8 pc of the Sun.  This ordered
component rotates slowly at a rate of $\sim 0.25^\circ$ per parsec.
Superimposed on this field is a turbulent component of $\sim 23^\circ$
\cite{Frisch:2012ismf2}.  Since the Ribbon arc can be centered
up to 16\deeg\ away from the ISMF direction outside of the heliosphere
according to the models
\cite{HeerikhuisenPogorelov:2011}, and the field rotates slowly, 
we consider the ISMF directions obtained from the polarization data to
be consistent with the direction of the ISMF that shapes the
heliosphere.

\begin{figure}[h!]
\begin{centering}
  \resizebox{17pc}{!}{\includegraphics{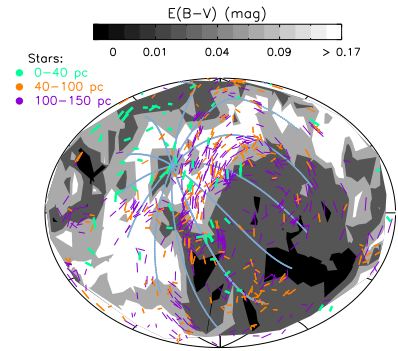}}
\end{centering}
\caption{The directions of the ISMF within 40 pc,
and polarized starlight, are shown projected against the
distribution of interstellar dust within 100 pc.  Starlight polarized
by interstellar dust traces the magnetic field direction.  The
blue-green lines trace the ISMF direction and intersect at a pole of the
ISMF that best fits the optical polarization data.  Measurements of
stellar color excesses, \ebv, provide the the dust distribution. The
white arc-like regions show the distribution of interstellar dust that
traces the boundary regions of Loop I within 100 pc.  The dark regions
in the third and fourth galactic quadrants correspond to the deficit
of local ISM associated with the Local Bubble, which also corresponds
to the regions with the brightest local UV and EUV radiation fluxes
because of low interstellar opacities. The data are plotted in an
aitoff projection, centered on the galactic center, and with longitude
increasing to the left.  See \cite{Frisch:2012ismf2} for more details
}.
\label{fig:loopI}
\end{figure}

Figure \ref{fig:loopI} shows the direction of the best-fitting local
ISMF obtained from the optical polarization data, compared to the
distribution of interstellar dust within 100 pc.  The large white
arc-like region maps the interstellar dust within 100 pc that is
associated with Loop I, a giant magnetic bubble that
appears to extend to the solar location (see below).  The very local
ISMF may be part of the Loop I magnetic field.

\section{Heliosphere Tracers of ISM}

 The relative Sun-interstellar velocity drives interstellar gas and
larger dust grains into the inner heliosphere, while the interstellar
magnetic field, ionized gas, and tiny charged grains are deflected
around the heliopause.  Neutral interstellar gas is ionized in the
heliosphere by charge-exchange (CEX) with solar wind plasma,
photoionization, and/or electron impact ionization, depending on the
species.  Charge exchange between the solar wind and interstellar
\HI\ creates ENAs that are mapped by IBEX, revealing the time-variable
heliosphere plasmas.  Ionized interstellar atoms are convected
outwards with the solar wind as a population of evolving pickup ions
(PUI) that seed the formation of anomalous cosmic rays (ACR).  All of
the abundant atoms that have significant neutral fractions in the
interstellar cloud around the heliosphere, e.g. H, He, N, O, Ne, and
Ar, have been either directly detected or observed in the form of PUIs
or ACRs \cite{Frisch:2011araa}.  When combined, these data show that
the LIC is a warm, low density, partially ionized interstellar cloud,
with the isotopic composition of the Sun.  It is flowing through the
heliosphere from a direction within 15\deeg\ of the galactic
center \cite{Frischetal:2009ibex,Frisch:2011araa}.

The penetration depth of neutral interstellar atoms into the
heliosphere differs between elements, depending mainly on the CEX
cross-section.  Helium and neon are on hyperbolic trajectories,
becoming ionized within 1 AU of the Sun so that surviving
atoms form the focusing tail, an elongated density
enhancement downwind of the Sun 
\cite{Moebius:2004he,Bochsler:2012ne}.

The parameters of the LIC at the heliosphere
boundaries have been reviewed in \cite{Frisch:2011araa}.
Densities of \HI, protons, and \HeI\ are, respectively, 0.19 \cc, 0.07
\cc, and  $0.015 \pm 0.003$ \cc.  The \HeI\ density is
based on Ulysses measurements up to several AU from the Sun, where the
gravitational distortion of particle trajectories into the downwind
focusing cone is minimized \cite{Moebius:2004he}.  Measurements of
interstellar \HeI\ at 1 AU by IBEX-LO show that ISM flows through the
heliosphere at a velocity of $23.2 \pm 0.3$ \kms\ and a temperature of
$6300 \pm 390$ K, with the flow directed towards galactic coordinates
of \glon,\glat$=185.25^\circ \pm 0.24^\circ, -12.03^\circ \pm
0.51^\circ$ (\cite{McComas:2012bow}, see Table 1 for ecliptic
coordinates).

In contrast, the density distribution of neutral interstellar hydrogen
is strongly modified by CEX inside of the heliosphere and in the inner
and outer heliosheath regions, and by radiation pressure. The complex
\HI\ source function, with contributions from primary and secondary
particle populations, temporal and latitude variations in the solar
L$\alpha$ flux that affect trajectories and backscattered emission,
and $\sim 50$\% H filtration, complicates efforts to extract an
accurate LIC velocity vector from the \HI\ backscattered data
\cite{QuemeraisIzmodenov:2002,Quemerais:1996tdot,PryorAjello:2003}.
Thus the most reliable measure of the LIC velocity is obtained from
data on interstellar
\HeI\ inside of the heliosphere.

\section{Changes in Interstellar Wind}

\subsubsection{Historical helium data}

The velocity vector of the flow of interstellar \HeI\ through the
heliosphere has been determined many times over the past 40 years.
The directions of the \HeI\ flow have been either determined from the
weak resonant scattering of solar 584\AA\ emission by interstellar
\HeI\ in the inner heliosphere, or from \emph{in situ} \HeI\
 measurements.

\begin{figure}[h!]
\begin{centering}
  \resizebox{17pc}{!}{\includegraphics{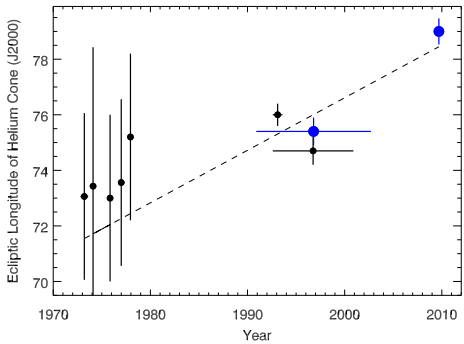}}
\end{centering}
\caption{Directions of the flow of interstellar \HeI\ from direct
detection of He atoms (after 1990), and the fluorescence of the EUV
584\AA\ line (20th century, see Table 1).  Although the earlier
results have been discounted \cite{Lallement:2004he} since they did
not agree with the results of the Ulysses GAS detector
\cite{Witte:2004,Moebius:2004he}, the most recent precise measurement
by IBEX \cite{McComas:2012bow} indicate that there may be a real
temporal variation in the direction of the ISM flowing into the
heliosphere. If so, a linear fit to the results plotted in this figure suggest that
the ecliptic longitude of the wind direction has varied at 
a rate of roughly $\delta \lambda \sim 0.19^\circ$ per
year.  The width of the horizontal bars for each point show the
earliest and latest times over which the data were collected.  The
large blue points are the IBEX and Ulysses \emph{in situ} values. }
\label{fig:he}
\end{figure}

These historical data indicate that the direction of the
\HeI\ flow vector may have varied slowly over time (Figure
\ref{fig:he}). The blue points represent directions determined 
from \emph{in situ} measurements of \HeI\ by Ulysses \cite{Witte:2004}
and IBEX \cite{McComas:2012bow}, while the remaining points indicate
584\AA\ backscattered values.  The
\HeI\ velocity, beginning and end times of the measurements, and
downwind ecliptic longitude of the velocity are listed in Table 1,
where most directions have been precessed to J2000 coordinates.

Interstellar helium enters the heliosphere with minimal filtration,
and trajectories are determined by the position, velocity, and
temperature of the particles, solar gravity, and ionization due to
photoionization and electron impact ionization within 1 AU
\cite{Moebius:2012isn,Bzowski:2012isn}. 
The details of the models used to obtain the
\HeI\ vectors vary between the different sets of data, particularly
for the older data where the models are less developed and solar flux
data were not reliable.

These data provide a basis for evaluating temporal variations in the
interstellar boundary conditions of the heliosphere.  Figure
\ref{fig:he} shows that the ecliptic longitude of the \HeI\ velocity
vector has increased with time over the past forty years (Table 1).  A
linear fit to the data suggests an average increase of $\sim
0.19^\circ$ per year in the ecliptic longitude of the He flow.  Over
the same interval the rotation of the ISMF at the Sun would be only
1/100th of an arcminute based on the nearby ordered component of the
ISMF shown by the polarization data found in \cite{Frisch:2012ismf2}.
At this rate, the direction of the LIC flow would change by 90\deeg\
over 500 years.  Depending on the behavior of the ISMF, the
heliosphere configuration should also vary significantly over
century-length timescales.

Assuming that the time-dependence of the \HeI\ direction is real, I
will now look at possible interstellar explanations for this
variation.

\subsubsection{Why the longitude change?}
It has been argued that the older data are inaccurate because of
instrumental uncertainties or poorly known ionization corrections.  In
an effort to make the older Prognoz 6 \HeI\ data
\cite{DalaudierBertaux:1984} consistent with  the Ulysses GAS
results \cite{Witte:2004}, Lallement et al. \cite{Lallement:2004he}
found that the two sets of results could be brought into agreement if
the instrumental noise was decoupled from the pre-launch laboratory
calibration and added as a variable into the analysis.  However this
type of argument can not explain the even steeper temporal variation
between the wind longitudes found by IBEX versus Ulysses (Figure
\ref{fig:he}).

One might speculate that the slow increase in ecliptic longitude of the
velocity is due to variations in solar conditions over the past 40
years.  At a given spatial position, a \HeI\ atom may have arrived
directly from the ISM, or may have arrived indirectly through close
passage by the Sun on the downwind side due to gravitational focusing
\cite{MuellerCohen:2012}.  An apparent shift in the direction could
occur if there were a systematic variation in the mix of direct and
indirect atoms that are detected, perhaps due to variations in the
ionization rates of the indirect atoms traveling close to the Sun
where losses from electron impact ionization and photoionization
dominate. This argument would not apply to the Ulysses value since
indirect atoms contributed only a third of the data used by Witte
(2004, \cite{Witte:2004}).  Sunspot activity levels vary between the
different sets of measurements, but no systematic trend is associated
with the data listed in the Table.

A more interesting possibility is that the direction of the ISM flow
through the heliosphere has varied with time, as suggested by the
historical measurements of the flow of neutral interstellar helium
atoms through the heliosphere.

\subsubsection{Cloud edges}
Cloud edges are interesting properties of the ISM that are difficult
to study, and the Sun is near a cloud edge \cite{Frisch:2011araa}.  It
is now well known that an absorption component at the LIC velocity is
not observed towards stars in the upwind direction.  The most
compelling example is 36 Oph, which is a star 6 pc away and $9^\circ$
from the heliosphere nose direction.  Towards this star, the
velocities of the interstellar
\FeII, \MgII, and \DI\ lines are $28.3 \pm 0.5$
\kms\ \cite{Wood36Oph:2000} in contrast to the $\sim 23.2$ \kms\ that is
predicted from the \HeI\ LIC velocity.  This indicates that for an
average LIC density of 0.19 \cc, the Sun is less than 20,000 AU (0.1
pc) from the LIC boundary.

The LIC belongs to an ISM flow (the cluster of local interstellar
clouds, CLIC) with an upwind direction in the local standard of rest
(LSR) that is within 15\deeg\ of the center of the S1 shell, directed
towards the center of the Loop I superbubble
\cite{Frisch:2011araa,Frisch_etal_2002}.
The properties of the edge of the LIC depend on the adjacent ISM.

The CLIC is embedded in the hot Local Bubble plasma.  Since neutral
gas fills 25\%--40\% of the sightlines towards nearby stars, or
less if clouds consist of randomly distributed uniform spherical
objects, the LIC may be bounded by million degree tenuous plasma.  For
this scenario, the LIC edge may consist either of a hot evaporative
conductive interface, or of a turbulent mixing layer
\cite{Frisch:2011araa}.  Both phenomena disrupt the velocities of
cloud edges.

The CLIC is decelerating.  This property is shown by the velocity
distribution of clouds in the flow, some of which are accelerated
towards the Sun, relative to the mean flow velocity, in both the
upwind and downwind directions
\cite{Frisch:2011araa,Frisch_etal_2002}.
In a decelerating flow, velocities in cloud boundary regions may be
disrupted by collisions between clouds that create density
enhancements similar to the Leo clouds discussed in Meyer et
al. \cite{Meyer:2012sss}, or by other instabilities.

Fluctuations in interstellar electron densities are identified over
scales of $\sim 10^4$ km through 10 AU by their effect on radio wave
propagation \cite{Armstrongetal:1995}. Radio scintillation data 
showing that electron scattering screens are present within 10 pc has
been interpreted to indicate the presence of local cloud collisions
\cite{Linsky_etal_2008}.

One piece of information that we have is that the change in the
flow direction is such that the flow is moving away from the direction
of the ISMF that shapes the heliosphere, which is towards
\elon,\elat$=221^\circ,39^\circ$ according to the center of the IBEX
Ribbon arc
\cite{McComas:2009sci,Funsten:2009sci}.  Note that the latitude
variations are smaller and less well constrained than the longitude
variations.  This would seem to indicate
that the ISM feeding the \HeI\ into the heliosphere has become less
ionized over time, since increased ionization would couple the flow
more tightly to the ISMF.  The decreasing-ionization scenario is
consistent with a local gradient in the interstellar radiation field
that is defined by the lower opacity and higher radiation fluxes found
in the galactic interval \glon$\sim 190^\circ - 360^\circ$ (Figure
\ref{fig:loopI},
\cite{Frisch:2012ismf2}.

\subsubsection{Turbulence}

Based on the trend for the
ecliptic longitude of the LIC velocity vector to increase with time,
it should be considered a real possibility that the Sun has sampled
turbulence in the LIC velocity over the past several decades.  

The average values for the temperature and turbulence of the LIC are
$7,500 \pm 1,300$ K and $1.62 \pm 0.75$ \kms, based on absorption
lines toward 19 stars behind the LIC according to Redfield et al.
\cite{RLIV:2008}.  Nonthermal turbulence is calculated by assuming that
the full-width-half-max of the absorption line is proportional to
$(V_\mathrm{thermal}^2 + V_\mathrm{turbulence}^2)^{1/2}$, where
$V_\mathrm{thermal}$ is the Doppler spread in thermal velocities
appropriate for the cloud temperature, and $ V_\mathrm{turbulence}$
represents non-thermal contributions to the velocity over the length
of the sightline.  Uncertainties in this type of analysis can
incorporate gradients in the velocity field, or even blended
absorption components, into the turbulence parameter depending on
spectral resolution and data quality (e.g. \cite{WeltyK:2001}).

If the average line width of the LIC is attributed to the ISM entering
the heliosphere, but with a temperature $\sim 6,300$ K as found from
the \HeI\ data, then the turbulent velocity would be 4.7 \kms.  There
is no information about the scale size of the turbulence in the LIC,
so according to these arguments, a plausible case can be made that
this nominal 4.6 \kms\ turbulent component samples turbulence in the
ISM between the Sun and the upwind edge of the LIC.  In such a case, a
systematic variation in the flow direction is possible.

\subsubsection{Filaments in the magnetoionic medium}

A gradient in the magnetic field strength between the heliosphere and
the edge of the LIC is also possible.  Turbulence in the galactic
magnetic field manifests itself in the magnetoionic ISM over large
spatial scales through structures that appear in maps of the polarized
radio continuum, but which are not visible in the radio continuum
intensity maps (for a review of the magnetoionic medium see
\cite{Haverkorn:2010magnetoionic}.)  Gaensler et al. \cite{Gaensler:2011mfturb} have
shown that these structures form a web of filaments that do not appear
to be supersonic.  Most interesting for the local ISM is that the
spatial gradient of the radio continuum polarization is such that the
polarization changes more rapidly for directions that are
perpendicular to the filamentary elongations.  If this is the case for
the LIC, and the LIC is elongated along the ISMF direction as given by
the center of the IBEX Ribbon arc, the gradient in the ISMF would make
an angle of $\sim 42^\circ$ with respect to the \HeI\ flow vector
(since the spherical angle between the ISW and ISMF is 48$^\circ$).  A
gradient in the ISMF could affect the heliosphere shape, and perhaps
the \HeI\ flow direction over time.

\subsubsection{Loop I superbubble}

A hand-waving explanation for the variations of the direction of
the interstellar \HeI\ flow through the heliosphere 
is that the LIC is turbulent 
because it is part of the Loop I superbubble.  Three epochs of star
formation in the nearby Sco-Cen Association formed Loop I
(e.g.\cite{Frisch:2011araa}).  Comparisons between optical and radio
synchrotron emission polarization data show that the ISMF is parallel
to the elongated filaments that form the observed shell-like structure
of Loop I (e.g. \cite{Heiles:1998lb}).  The brightest region of Loop I
in the 1.8 MHz radio continuum is the North Polar Spur, 
where the ISM density increases continuously from the Sun
out to $\sim 100$ pc
\cite{Santosetal:2010,PlanetPol:2010,Frisch:2012ismf2}. The local
ordered ISMF discussed above is directed towards the North Polar Spur.

Three sets of data suggest that the Sun is in a fragment of
the shell of the Loop I superbubble: (1) The Loop I model of Wolleben
\cite{Wolleben:2007,Frisch:2010s1} identifies two radio continuum
shells (called ``S1'' and ``S2'') at different distances and
locations.  The Sun is located in the rim of the S1 shell according to
his model, a result that agrees with prior estimates of the Loop I
configuration \cite{Frisch:1996}.  (2) Interstellar clouds within
about 20 pc of the Sun, e.g. the CLIC, have a bulk velocity in the LSR
that is within 15\deeg\ of the center of the S1 shell, suggesting
a dynamical expanding shell configuration
\cite{Frisch:2010s1,Frisch:2011araa}.
(3) The local ISMF directions found from the center of the IBEX Ribbon
arc, and from the interstellar polarization measurements, both have
angles $\sim 76^\circ$ away from the bulk local ISM flow, and $\sim
64^\circ$ from the S1 center as defined by Wolleben.

In conclusion, there are many astronomical
justifications for the hypothesis that the very local ISM varies over
spatial scales comparable to the heliosphere dimensions.  After all,
the heliosphere, as well as other stars
\cite{RansomWolleben:2010PNtail}, disrupt the ISM and the ISMF over
thousand-AU spatial scales.  A detailed study of the historical \HeI\
data is needed to evaluate the possibility that the direction of the
flow of interstellar gas through the heliosphere has varied over the
space age.


\begin{theacknowledgments}
This work has been supported by the IBEX mission as part
of the NASA Explorer Program.
\end{theacknowledgments}


\begin{thebibliography}{58}
\expandafter\ifx\csname natexlab\endcsname\relax\def\natexlab#1{#1}\fi
\providecommand{\enquote}[1]{``#1''}
\expandafter\ifx\csname url\endcsname\relax
  \def\url#1{\texttt{#1}}\fi
\expandafter\ifx\csname urlprefix\endcsname\relax\def\urlprefix{URL }\fi
\providecommand{\eprint}[2][]{\url{#2}}


\bibitem[{Frisch} and {Mueller}(2011)]{FrischMueller:2011ssr}
P.~C. {Frisch}, and H.-R. {Mueller}, \emph{\ssr} p.130 (2011).

\bibitem[{Davis}(1955)]{Davis:1955}
L.~{Davis}, \emph{Physical Review} \textbf{100}, 1440--1444 (1955).

\bibitem[{Holzer}(1989)]{Holzer:1989}
T.~E. {Holzer}, \emph{\araa} \textbf{27}, 199--234 (1989).

\bibitem[{Stone} et~al.(2008)]{Stoneetal:2008nature}
E.~C. {Stone}, A.~C. {Cummings}, F.~B. {McDonald}, B.~C. {Heikkila}, N.~{Lal},
  and W.~R. {Webber}, \emph{\nat} \textbf{454}, 71--74 (2008).

\bibitem[{Frisch} et~al.(2011)]{Frisch:2011araa}
P.~C. {Frisch}, S.~Redfield, and J.~Slavin, \emph{\araa} \textbf{49} (2011).

\bibitem[{McComas} et~al.(2009)]{McComas:2009sci}
D.~J. {McComas}, F.~{Allegrini}, P.~{Bochsler}, et~al.,
\emph{Science} \textbf{326},
  959 (2009).

\bibitem[{Lazarian} and {Desiati}(2010)]{LazarianDesiati:2010}
A.~{Lazarian}, and P.~{Desiati}, \emph{\apj} \textbf{722}, 188--196 (2010).

\bibitem[{Schwadron} et~al.(2012)]{Schwadron:2012gcr}
N.~A. {Schwadron}, F.~C. {Adams}, B.~{Dingus}, 
et~al., \emph{in preparation} (2012).


\bibitem[{Saul} et~al.(2012)]{GALFA:2012sss}
D.~R. {Saul}, J.~E.~G. {Peek}, J.~{Grcevich}, 
  et~al., \emph{\apj} \textbf{758}, 44 (2012).

\bibitem[{Meyer} et~al.(2012)]{Meyer:2012sss}
D.~M. {Meyer}, J.~T. {Lauroesch}, J.~E.~G. {Peek}, and C.~{Heiles}, \emph{\apj}
  \textbf{752}, 119 (2012).

\bibitem[{Pogorelov} et~al.(2009)]{Pogorelovetal:2009ibexribbonasymmetries}
N.~V. {Pogorelov}, J.~{Heerikhuisen}, J.~J. {Mitchell}, I.~H. {Cairns}, and
  G.~P. {Zank}, \emph{\apjl} \textbf{695}, L31--L34 (2009).

\bibitem[{Slavin} and {Frisch}(2008)]{SlavinFrisch:2008}
J.~D. {Slavin}, and P.~C. {Frisch}, \emph{\aap} \textbf{491}, 53--68 (2008).

\bibitem[{Schwadron} et~al.(2009)]{Schwadron:2009sci}
N.~A. {Schwadron}, M.~{Bzowski}, G.~B. {Crew}, M.
  et~al., \emph{Science}
  \textbf{326}, 966 (2009).

\bibitem[{Schwadron} et~al.(2011)]{Schwadronetal:2011sep}
N.~A. {Schwadron}, F.~{Allegrini}, M.~{Bzowski}, 
et~al., \emph{\apj} \textbf{731}, 56--77 (2011).

\bibitem[{Frisch}(2012)]{Frisch:2012howlocal}
P.~C. {Frisch}, \enquote{{How local is the local interstellar magnetic
  field?},} in \emph{\aipcs}, edited by J.~{Heerikhuisen}, G.~{Li},
  N.~{Pogorelov}, and G.~{Zank}, 2012, vol. 1436 of \emph{\aipcs}, 
  295--301.

\bibitem[{Heerikhuisen} et~al.(2010)]{Heerikhuisen:2010ribbon}
J.~{Heerikhuisen}, N.~V. {Pogorelov}, G.~P. {Zank}, G.~B. {Crew},
  et~al., \emph{\apjl} \textbf{708}, L126--L130
  (2010).

\bibitem[{Heerikhuisen} and {Pogorelov}(2011)]{HeerikhuisenPogorelov:2011}
J.~{Heerikhuisen}, and N.~V. {Pogorelov}, \emph{\apj} \textbf{738}, 29
  (2011).

\bibitem[{Ratkiewicz} et~al.(2012)]{Ratkiewicz:2012ribbon}
R.~{Ratkiewicz}, M.~{Strumik}, and J.~{Grygorczuk}, \emph{\apj} \textbf{756}, 3
  (2012).

\bibitem[{Chalov} et~al.(2010)]{Chalov:2010ribbon}
S.~V. {Chalov}, D.~B. {Alexashov}, D.~{McComas}, et~al.,
 \emph{\apjl} \textbf{716}, L99--L102 (2010).

\bibitem[{Frisch} and {McComas}(2010)]{FrischMcComas:2010}
P.~C. {Frisch}, and D.~J. {McComas}, \emph{\ssr}  (2010).

\bibitem[{Frisch} et~al.(2012)]{Frisch:2012ismf2}
P.~C. {Frisch}, B.~{Andersson}, A.~{Berdyugin}, et~al.,
  \emph{\apj}, \textbf{760},
  160--178 (2012).

\bibitem[{Andersson}(2012)]{Andersson:2012rev}
B.-G. {Andersson}, \emph{ArXiv e-prints/astro-ph:1208.4393}  (2012).

\bibitem[{Frisch} et~al.(2009)]{Frischetal:2009ibex}
P.~C. {Frisch}, M.~Bzowski, E.~{Gr{\"u}n}, 
  et~al., \emph{\ssr} \textbf{146}, 235--273 (2009).

\bibitem[{M{\"o}bius} et~al.(2004)]{Moebius:2004he}
E.~{M{\"o}bius}, M.~{Bzowski}, S.~{Chalov}, et~al.,
  \emph{\aap} \textbf{426}, 897--907 (2004).

\bibitem[{McComas} et~al.(2012)]{McComas:2012bow}
D.~J. {McComas}, D.~{Alexashov}, M.~{Bzowski}, et al.,
  \emph{Science} \textbf{336}, 1291 (2012).

\bibitem[{Bochsler} et~al.(2012)]{Bochsler:2012ne}
P.~{Bochsler}, L.~{Petersen}, E.~{M{\"o}bius}, et al.,
  \emph{\apjs} \textbf{198}, 13 (2012).

\bibitem[{Qu{\'e}merais} and {Izmodenov}(2002)]{QuemeraisIzmodenov:2002}
E.~{Qu{\'e}merais}, and V.~{Izmodenov}, \emph{\aap} \textbf{396}, 269--281
  (2002).

\bibitem[{Quemerais} et~al.(1996)]{Quemerais:1996tdot}
E.~{Quemerais}, Y.~G. {Malama}, W.~R. {Sandel}, et all,
\emph{\aap} \textbf{308}, 279--289 (1996).

\bibitem[{Pryor} et~al.(2003)]{PryorAjello:2003}
W.~R. {Pryor}, J.~M. {Ajello}, D.~J. {McComas}, et al.
  \emph{Journal of Geophysical Research (Space Physics)}
  \textbf{108}, 9--1 (2003).

\bibitem[{Witte}(2004)]{Witte:2004}
M.~{Witte}, \emph{\aap} \textbf{426}, 835--844 (2004).

\bibitem[{M{\"o}bius} et~al.(2012)]{Moebius:2012isn}
E.~{M{\"o}bius}, P.~{Bochsler}, M.~{Bzowski}, et al.
  \emph{\apjs} \textbf{198}, 11 (2012).

\bibitem[{Bzowski} et~al.(2012)]{Bzowski:2012isn}
M.~{Bzowski}, M.~A. {Kubiak}, E.~{M{\"o}bius}, et al.,
  \emph{\apjs} \textbf{198}, 12 (2012).

\bibitem[{Dalaudier} et~al.(1984)]{DalaudierBertaux:1984}
F.~{Dalaudier}, J.~L. {Bertaux}, V.~G. {Kurt}, and E.~N. {Mironova},
  \emph{\aap} \textbf{134}, 171--184 (1984).

\bibitem[{Lallement} et~al.(2004)]{Lallement:2004he}
R.~{Lallement}, J.~C. {Raymond}, J.~{Vallerga}, et al.,
\emph{\aap} \textbf{426}, 875--884 (2004).

\bibitem[{M{\"u}ller} and {Cohen}(2012)]{MuellerCohen:2012}
H.-R. {M{\"u}ller}, and J.~H. {Cohen}, \enquote{{Primary neutral helium in the
  heliosphere},} in \emph{\aipcs}, edited by J.~{Heerikhuisen}, G.~{Li},
  N.~{Pogorelov}, and G.~{Zank}, 2012, vol. 1436 of \emph{\aipcs}, 
  233--238.

\bibitem[{Wood} et~al.(2000)]{Wood36Oph:2000}
B.~E. {Wood}, J.~L. {Linsky}, and G.~P. {Zank}, \emph{\apj} \textbf{537},
  304--311 (2000).

\bibitem[{Frisch} et~al.(2002)]{Frisch_etal_2002}
P.~C. {Frisch}, L.~{Grodnicki}, and D.~E. {Welty}, \emph{\apj} \textbf{574},
  834--846 (2002).

\bibitem[{Armstrong} et~al.(1995)]{Armstrongetal:1995}
J.~W. {Armstrong}, B.~J. {Rickett}, and S.~R. {Spangler}, \emph{\apj}
  \textbf{443}, 209--221 (1995).

\bibitem[{Linsky} et~al.(2008)]{Linsky_etal_2008}
J.~L. {Linsky}, B.~J. {Rickett}, and S.~{Redfield}, \emph{\apj} \textbf{675},
  413--419 (2008).

\bibitem[{Funsten} et~al.(2009)]{Funsten:2009sci}
H.~O. {Funsten}, F.~{Allegrini}, G.~B. {Crew}, et al.,
  \emph{Science} \textbf{326}, 964--967 (2009).


\bibitem[{Weller} and {Meier}(1974)]{WellerMeier:1974}
C.~S. {Weller}, and R.~R. {Meier}, \emph{\apj} \textbf{193}, 471--476 (1974).

\bibitem[{Ajello} et~al.(1979)]{Ajello:1979}
J.~M. {Ajello}, N.~{Witt}, and P.~W. {Blum}, \emph{\aap} \textbf{73}, 260--271
  (1979).

\bibitem[{Weller} and {Meier}(1979)]{WellerMeier:1979}
C.~S. {Weller}, and R.~R. {Meier}, \emph{\apj} \textbf{227}, 816--823 (1979).

\bibitem[{Weller} and {Meier}(1981)]{WellerMeier:1981}
C.~S. {Weller}, and R.~R. {Meier}, \emph{\apj} \textbf{246}, 386--393 (1981).

\bibitem[{Flynn} et~al.(1998)]{FlynnVallerga:1998}
B.~{Flynn}, J.~{Vallerga}, F.~{Dalaudier}, and G.~R. {Gladstone}, \emph{\jgr}
  \textbf{103}, 6483 (1998).

\bibitem[{Vallerga} et~al.(2004)]{Vallergaetal:2004}
J.~{Vallerga}, R.~{Lallement}, M.~{Lemoine}, F.~{Dalaudier}, and D.~{McMullin},
  \emph{\aap} \textbf{426}, 855--865 (2004).

\bibitem[{Adams} and {Frisch}(1977)]{AdamsFrisch:1977}
T.~F. {Adams}, and P.~C. {Frisch}, \emph{\apj} \textbf{212}, 300--308 (1977).


\bibitem[{Redfield} and {Linsky}(2008)]{RLIV:2008}
S.~{Redfield}, and J.~L. {Linsky}, \emph{\apj} \textbf{673}, 283--314 (2008).

\bibitem[{Welty} and {Hobbs}(2001)]{WeltyK:2001}
D.~E. {Welty}, and L.~M. {Hobbs}, \emph{\apjs} \textbf{133}, 345--393 (2001).

\bibitem[{Haverkorn}(2010)]{Haverkorn:2010magnetoionic}
M.~{Haverkorn}, \enquote{{The Magneto-Ionic Medium in the Milky Way},} in
  \emph{Astr.Soc.Pac.Conf.Ser.}, edited by
  R.~{Kothes}, T.~L. {Landecker}, and A.~G. {Willis}, 2010, vol. 438 
	p. 249, \eprint{1012.3755}.

\bibitem[{Gaensler} et~al.(2011)]{Gaensler:2011mfturb}
B.~M. {Gaensler}, M.~{Haverkorn}, B.~{Burkhart}, et al.,
  \emph{\nat} \textbf{478}, 214--217 (2011).

\bibitem[{Heiles}(1998)]{Heiles:1998lb}
C.~{Heiles}, \enquote{{The Magnetic Field Near the Local Bubble},} in \emph{IAU
  Colloq. 166: The Local Bubble and Beyond}, edited by D.~{Breitschwerdt},
  M.~J. {Freyberg}, and J.~{Truemper}, 1998, vol. 506 of \emph{Lecture Notes in
  Physics, Berlin Springer Verlag}, 229--238.

\bibitem[{Santos} et~al.(2011)]{Santosetal:2010}
F.~P. {Santos}, W.~{Corradi}, and W.~{Reis}, \emph{\apj} \textbf{728}, 104
  (2011).

\bibitem[{Bailey} et~al.(2010)]{PlanetPol:2010}
J.~{Bailey}, P.~W. {Lucas}, and J.~H. {Hough}, \emph{\mnras} \textbf{405}, 2570--2578 (2010).

\bibitem[{Wolleben}(2007)]{Wolleben:2007}
M.~{Wolleben}, \emph{\apj} \textbf{664}, 349--356 (2007).

\bibitem[{Frisch}(2010)]{Frisch:2010s1}
P.~C. {Frisch}, \emph{\apj} \textbf{714}, 1679--1688 (2010).

\bibitem[{Frisch}(1996)]{Frisch:1996}
P.~C. {Frisch}, \emph{\ssr} \textbf{78}, 213--222 (1996).

\bibitem[{Ransom} et~al.(2010)]{RansomWolleben:2010PNtail}
R.~R. {Ransom}, R.~{Kothes}, M.~{Wolleben}, and T.~L. {Landecker}, \emph{\apj}
  \textbf{724}, 946--956 (2010).

\end{thebibliography}

\end{document}